\documentclass{IEEEtran4PSCC}

\usepackage{siunitx}
\usepackage{comment}
\usepackage{booktabs}
\usepackage{makecell}
\usepackage{cite}

\usepackage{graphicx}
\usepackage[cmex10]{amsmath}

\usepackage{multicol}

\begin{document}
%

\title{Modeling and Control of Hybrid Distribution Transformers for  Simultaneous Grid Services}

\IEEEoverridecommandlockouts

\author{
Martin Doff-Sotta$^{*}$\thanks{$^{*}$Department of Engineering Science, University of Oxford, Oxford, UK.},  Florian Cech$^{\dagger}$\thanks{$^{\dagger}$ Ionate Ltd, London, UK. },  Rishabh Manjunatha$^{\dagger}$, \\ Costantino Citro$^{\dagger}$,  Matthew Williams$^{\dagger}$, and Thomas Morstyn$^{*}$
}


\maketitle





\thanksto{\noindent Submitted to the 24th Power Systems Computation Conference (PSCC 2026).}

\begin{abstract}Hybrid distribution transformers (HDTs) integrate conventional transformers with partially rated power electronic converters to improve power quality, enable advanced ancillary services and increase penetration of renewable energy sources in the national power grid.  In this paper, we present an averaged mathematical model of a three-phase HDT equipped with two back-to-back voltage source converters connected in a series-shunt configuration.  Cascaded PI controllers are designed in the synchronously rotating dq0 reference frame to regulate load voltage,  compensate reactive power, achieve grid frequency regulation,  and perform load phase balancing. Simulation results implemented in Python confirm that these simple yet effective control mechanisms allow HDTs to offer simultaneous grid services without introducing complexity. The complete model, control architecture, and implementation steps are detailed, enabling further validation and adoption. \end{abstract}

\begin{IEEEkeywords} Ancillary Services,  Frequency Regulation,  Hybrid Distribution Transformers,  Phase Balancing,  Power Factor Correction. \end{IEEEkeywords}

\section{Introduction}
The increasing penetration of distributed renewable energy resources and global electrification are creating unprecedented challenges for distribution network stability, power quality and flexibility. Conventional transformers offer no active control capabilities, making them inadequate for providing ancillary grid services.  By contrast,  hybrid distribution transformers (HDTs), which combine conventional transformers with partially rated power electronics, offer a promising path toward grid modernisation, enabling functions like voltage regulation, power factor correction, and frequency response \cite{burkard2015evaluation, bala2012hybrid}.

Compared to solid state transformers (SSTs), HDTs offer several engineering and economic advantages. The power electronic stage typically represents only 10–20\% of the total power rating, significantly reducing cost, complexity, and semiconductor stress \cite{bala2012hybrid, carreno2021configurations}.  Recent studies have also begun to quantify the economic value of HDTs, showing their potential to increase the export of power produced by distributed generators and reduce overall system costs when deployed strategically across distribution networks \cite{samuel}. 

HDTs have demonstrated capabilities in a wide range of grid services including power factor correction, harmonic mitigation, flicker suppression, phase shifting, and voltage regulation \cite{bala2012hybrid, carreno2021configurations, prystupczuk2022system}. More recent research has shown that HDTs can also support fault-tolerant operation by reverting to passive mode when the converter fails \cite{bala2012hybrid}, making them well suited for deployment in feeders with sensitive or bottleneck loads, and in long rural distribution lines.

In terms of control strategies, Proportional-Integral (PI) controllers are commonly used for local control of voltage and reactive power, while more advanced approaches rely on model predictive control (MPC) to coordinate multiple HDTs across a network \cite{kou2019decentralized, gao2021optimal}. Decentralised MPC frameworks are particularly attractive in active distribution networks, where local controllers use only local measurements and treat inter-device interactions as bounded disturbances, thereby avoiding the scalability issues and communication overhead associated with centralised control \cite{kou2019decentralized}.

Despite this progress, several gaps remain in the current literature. Most HDT studies assume simplified models or focus on steady-state performance, whereas there is a growing need for dynamic simulation and control frameworks that can handle realistic conditions such as unbalanced load and frequency deviations. Furthermore, while many existing HDT controllers target voltage regulation or power quality, relatively few examine frequency support, phase balancing, or the coordinated delivery of multiple ancillary services. 

This paper addresses these gaps by introducing an averaged three-phase model of an HDT based on first principles and demonstrating how simple decentralised control mechanisms can provide a wide range of ancillary services, including voltage regulation, power factor correction, frequency regulation, and phase balancing.  By contrast to existing approaches relying on complex control strategies~\cite{gao2021optimal},  the proposed PI control structure is both certifiable and implementable on standard low-cost microcontrollers. Moreover, whereas most prior studies focus on a single ancillary service,  we show that the proposed HDT topology enables simultaneous delivery of multiple services within a single device with minimal service cross-couplings. Finally,  we propose a series of scenarios reflecting realistic grid conditions, including unbalanced loads, grid voltage disturbances and varying grid frequency. 

The paper is organised as follows.  In Section \ref{sec:modeling}, we present a state-space model for an HDT equipped with two back-to-back voltage source controllers (VSCs) in a series-shunt topology.  Each converter is independently controlled using inner control loops described in Section \ref{sec:control}. The control structure is modular: the grid-side VSC regulates load voltage, while the load-side VSC manages shunt current injection and DC-link capacitor voltage stability.  Outer control loops are then proposed to handle sag / swell grid voltage rejection, reactive power compensation, frequency regulation, and load phase balancing.  Finally, in Section \ref{sec:results}, we demonstrate the capability of the HDT to deliver grid ancillary services  across various case studies. 

\section{Modeling}
\label{sec:modeling}
\begin{figure}[!ht]
    \centering
    \includegraphics[width=0.5\textwidth]{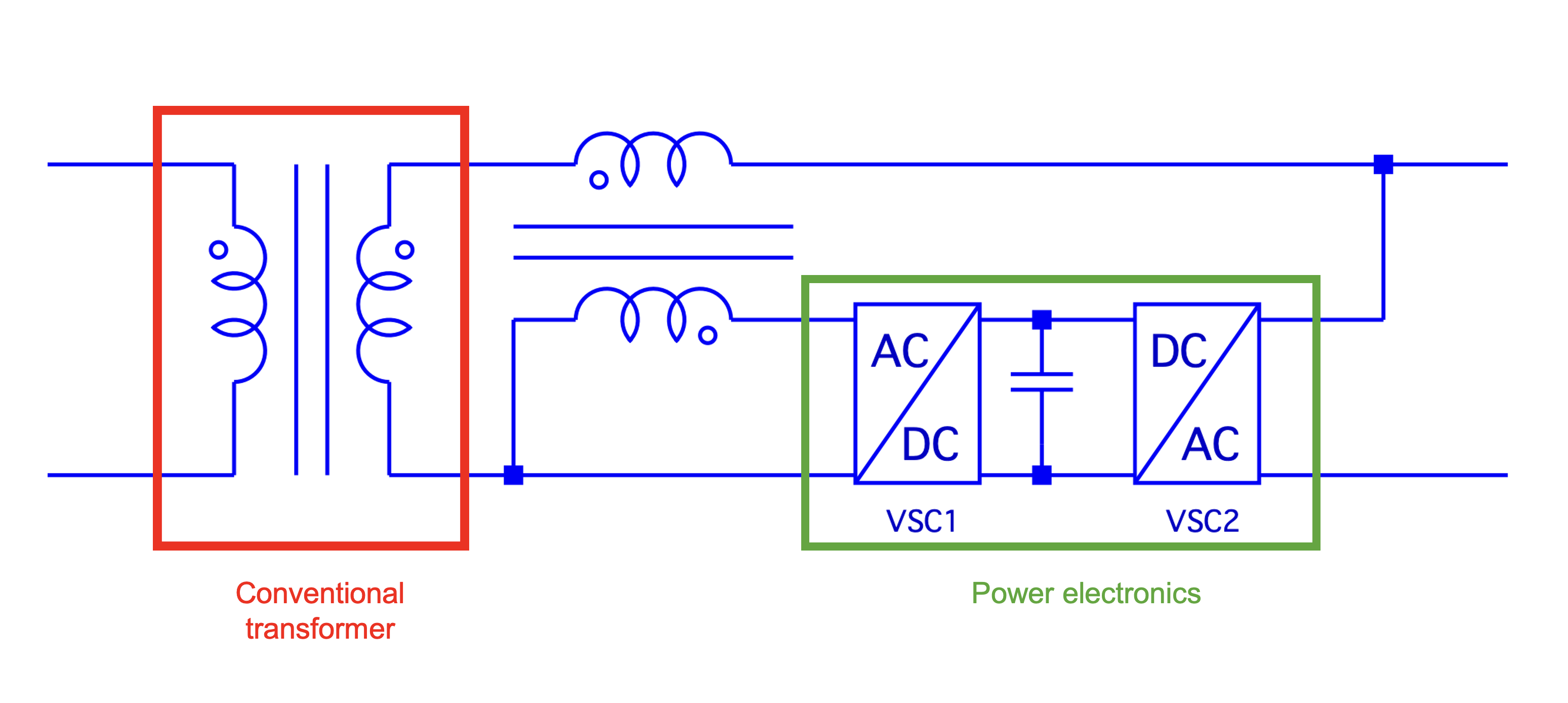} 
    \caption{HDT series-shunt topology.}
    \label{fig:topology}
\end{figure}

\begin{figure*}[!h]
    \centering
    \includegraphics[trim={3.5cm 15.5cm 4.5cm 11cm},clip, width=1\textwidth]{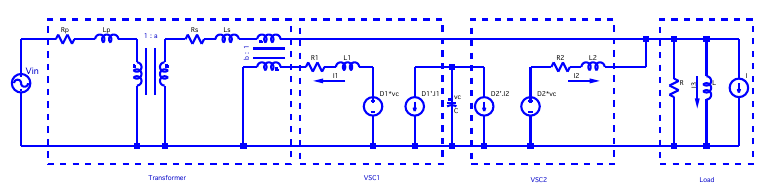}
    \caption{HDT averaged circuit.}
    \label{fig:diagram}
\end{figure*}

We consider an HDT with a bidirectional AC-DC-AC three-phase two-level converter connected in a series-shunt configuration, see Figure~\ref{fig:topology}. The two back-to-back VSCs are denoted by VSC1 (grid side) and VSC2 (load side). The dynamics of the power electronics is averaged over a switching period and the load is modeled as a three-phase parallel RL impedance with constant current source.  Figure~\ref{fig:diagram} shows an equivalent averaged circuit of the HDT.  It is worth noting that the HDT topology considered here closely resembles that of a Unified Power Quality Conditioner (UPQC) \cite{khadkikar2011enhancing}, suggesting that the control strategies and functionalities developed here may also be applicable to UPQC systems.

Referring to Figure \ref{fig:diagram}, $I_1, I_2, I_3$, are the 3-phase currents exiting VSC1, VSC2 and the load inductor (respectively), $v_C$ is the DC-link capacitor voltage, $V_\text{in}$ is the grid-side input voltage, $I$ is the 3-phase constant load current 
(where $I > 0$ corresponds to current absorption and $I < 0$ to current injection by the load) and $D_1$, $D_2$ are the 3-phase duty cycles of VSC1 and VSC2 (respectively). We also denote by $1/\alpha$ and $\beta$ the turn ratios of the transformers. 

From Kirchhoff's laws, we obtain the following governing equations for the different branches of the circuit

\begin{gather}
    \label{eq:I1}
    \bar{L}_1 \dot{I}_1  = -(\bar{R}_1 + \beta^2 R)I_1 + \beta R I_2 - \beta R I_3 \\ \nonumber - \beta R I - \alpha \beta V_\text{in} + D_1 v_C , \\
    \label{eq:I2}
     L_2 \dot{I}_2 = \beta R I_1 - \bar{R}_2 I_2 + R I_3 + R I  + D_2 v_C , \\
      \label{eq:I3}
     L\dot{I}_3 = -\beta R I_1 + R I_2 - R I_3 - R I, \\
     \label{eq:vc}
    C \dot{v}_C = -D_1^\top I_1   -D_2^\top I_2,
\end{gather}
where for compactness we defined  $\bar{L}_1 = L_1 + \beta^2(L_s + \alpha^2 L_p)$, $\bar{R}_1 = R_1 + \beta^2( R_s + \alpha^2 R_p)$, $\bar{R}_2 = R_2 + R$.   The load resistance $R$ and the load inductance $L$ are assumed to be matrices so that phase unbalance conditions can be analysed later.

\section{Control}
\label{sec:control}

\subsection{Inner control loop}

In this section, we detail the inner control loops of the back-to-back converters VSC1 and VSC2. Each converter is governed by an independent controller. VSC1 regulates the voltage on the load side,  while VSC2 tracks current setpoints to control the shunt current injection and stabilise the DC-link voltage.  A PI control structure is used to generate the duty cycles of the VSCs. 

\subsubsection{Load voltage control} The duty cycle of VSC1 is computed to track a reference voltage $V^*_{\text{dq0}} = [v^*\, 0 \, 0]^\top$ 

\begin{multline}
    \label{eq:d1}
    D_1 = \pi_{[-1;1]} ( T_{\omega}^{-1} ( K_{p, 1}( T_{\omega} V- V^*_{\text{dq0}}) \\+ K_{i, 1} \int ( T_{\omega}  V - V^*_{\text{dq0}}) \, \mathrm{d}t ) ), 
\end{multline}
where  $\pi_{[a; b]}(x) = \min(\max(x, a), b)$,  $V = - \beta R I_1 + R I_2 - R I_3 - R I $ is the voltage at the load terminals and $T_{\omega}$ is the abc- to dq0-frame transformation matrix: 
\[
T_{\omega} = {\frac{{2}}{{3}}} \begin{bmatrix}
    \sin{\omega t} & \sin{(\omega t-\frac{2\pi}{3})} & \sin{(\omega t+\frac{2\pi}{3})}\\
    \cos{\omega t} & \cos{(\omega t-\frac{2\pi}{3})} & \cos{(\omega t+\frac{2\pi}{3})}\\
    \frac{1}{{2}} & \frac{1}{{2}} & \frac{1}{{2}}
\end{bmatrix},
\]
such that $X_{dq0} = T_{\omega} X$ and $\omega = 2\pi f$ where $f$ is the grid frequency. 

\subsubsection{Shunt current control} The duty cycle of VSC2 is computed to track the reference currents in dq0 frame  $I^*_{2\text{dq0}}~=~[i^*_{2d} \, i^*_{2q} \, i^*_{20}]^\top$: 
\begin{multline}
    \label{eq:d2}
    D_2 = \pi_{[-1;1]} (T_{\omega}^{-1}( K_{p, 2}( I^*_{2\text{dq0}} - T_{\omega} I_2) \\+ K_{i, 2} \int (I^*_{2\text{dq0}} - T_{\omega} I_2)\, \mathrm{d}t)).
\end{multline}

\subsection{Outer control loop}

The outer control loop is responsible for generating the reference voltage $V^*_{\text{dq0}}$  and current  $I^*_{2\text{dq0}}$ in dq0- frame to achieve various ancillary services. 

\subsubsection{DC-link capacitor voltage control} For VSC2 to maintain the DC-link capacitor voltage within its operating range,   $i^*_{2d}$  is generated by the following PI controller 
\begin{equation}
    \label{eq:dc}
    i^*_{2d} =  K_{p, 3} (v_C - v_C^*) + K_{i, 3} \int (v_C - v_C^*)  \, \mathrm{d}t,
\end{equation}
where $v_C^*$ is the reference DC-link voltage. 

\subsubsection{Power factor correction} The goal of power factor correction is to ensure that the apparent power drawn by the grid-side interface of the HDT has a unity power factor. To achieve this, we define the augmented load as the combination of the physical load and the shunt converter (load + VSC2). The reactive power seen by the grid is then $\bar{Q} = Q - Q_{\text{sh}}$, where $Q$ is the load reactive power and $Q_{\text{sh}}$ is the shunt compensation. The VSC2 injects a q-axis current $i_{2q}$ to cancel $\bar{Q}$ using a feedforward + PI feedback structure. The controller is defined as:
\begin{equation}
\label{eq:pf_corr}
i_{2q}^* = i_{3q} + K_{p, 4} (\bar{Q}^* - \bar{Q}) + K_{i, 4} \int (\bar{Q}^* - \bar{Q})  \, \mathrm{d}t,
\end{equation}
where $i_{3q}$ is the measured q-axis component of the load inductor current, and $\bar{Q}^* = 0$ is the reactive power setpoint. This ensures the augmented load draws purely active power, achieving unity power factor at the point of common coupling.

\subsubsection{Frequency regulation}

The HDT provides frequency regulation by modulating its net active power injection based on grid frequency deviations.  A droop-based control scheme is adopted, whereby the desired power injection is computed from frequency deviations.   First define the frequency deviation $\Delta f = f - f_0$,  where $f$ is the measured grid frequency, $f_0 = 50 \si{Hz}$ is the nominal frequency.  Based on UK National Energy System Operator (NESO) dynamic regulation\footnote{NESO defines dynamic regulation as frequency response services that automatically adjust active power output in order to contain grid frequency within operational limits $\pm 0.2 \si{Hz}$ from the nominal frequency $f_0$.} response service \cite{NESO},  we define the droop curve coefficient $K_f(\Delta f)$ as follows 
\[
K_f(\Delta f) =
\begin{cases}
(P_{\text{max}} - P_0)/{\Delta f_{\text{max}}}, & \Delta f \geq 0, \\
(P_0 - P_{\text{min}})/{\Delta f_{\text{max}}}, & \Delta f < 0,
\end{cases}
\]
where $\Delta f_{\text{max}}$ is the maximum frequency deviation, $P_0$ is the nominal active power (corresponding to nominal load voltage and load impedance\footnote{We therefore assume that the load impedance is fixed or that we can forecast its value  and adapt the nominal power accordingly. This assumption is a limitation of our approach and will be relaxed in future work.}), and $P_{\text{min}}$ and $P_{\text{max}}$ are bounds corresponding to ±10\% voltage deviations from the nominal load voltage (beyond which performance degrades significantly). 

The desired power injection $P^*$ is then computed according to the piecewise-linear droop curve
\begin{equation}
\label{eq:freq_P}
P^* =   \pi_{[P_{\text{min}}; P_{\text{max}}]} (P_0 + K_f(\Delta f) \Delta f).
\end{equation}
The resulting reference power $P^*$ is converted into a  load voltage setpoint $v^*$ to be tracked by the inner control loop

\begin{equation}
\label{eq:freq_v}
v^* = v_0 + K_{p, 5} (P^* - P) + K_{i, 5} \int (P^* - P)  \, \mathrm{d}t,
\end{equation}
where $P$ is the load active power and $v_0$ is the nominal load voltage.  By modulating its voltage reference based on frequency deviations, the HDT delivers or absorbs active power in real time. This control strategy demonstrates the feasibility of active power bidding in the day-ahead frequency response market.


\subsubsection{Phase balancing} 

To ensure that the HDT mitigates unbalanced three-phase loads, a phase balancing controller is implemented in the synchronously rotating dq0 reference frame. The objective is to regulate the $q$-axis and zero-sequence components of the grid current to zero and stabilise the $d$-axis component to a constant reference value,  such that the load as seen from the grid appears balanced. 

Let $I_ \beta = - \beta I_1 $ denote the current injected by the secondary of the grid-side transformer and $I_{ \beta\text{dq0}}^{*} = [i_{ \beta\text{d}}^{*} \, 0 \, 0]^\top$ be the reference grid current in the  dq0 reference frame.  The controller estimates $i_{ \beta\text{d}}^{*}$ using the rolling RMS of past $d$-axis current measurements over one cycle. This ensures robustness to variations in the load. A PI controller is then used to compute the corrective current to inject through VSC2:

\begin{equation}
\label{eq:balancing0}
I_{2\text{dq0}}^\prime = {K}_{p, 6}  ( T_{\omega} I_\beta - I_{ \beta\text{dq0}}^{*}) + {K}_{i, 6} \int ( T_{\omega} I_\beta - I_{ \beta\text{dq0}}^{*})  \, \mathrm{d}t. 
\end{equation}

Finally, the $d$-axis component is adjusted to preserve the DC-link voltage control objective:
\begin{equation}
\label{eq:balancing}
I_{2\text{dq0}}^* =I_{2\text{dq0}}^\prime + \gamma \begin{bmatrix} i^*_{2d} \\ 0 \\ 0 \end{bmatrix}
\end{equation}
where $i^*_{2d}$ is the original reference for the $d$-axis current computed in equation \eqref{eq:dc} to achieve DC-link regulation, and $\gamma$ is a gain defining a tradeoff between DC-link voltage stability and load phase balancing.

\section{Numerical Experiments and Validation}
\label{sec:results}

The proposed HDT model and control architecture are validated using a Python-based simulation environment. Four case studies are conducted to assess the HDT ability to deliver a range of ancillary services. \textit{i) Load voltage regulation:} the HDT compensates for grid voltage sag and swell by adjusting the injected voltage through VSC1. The controller restores the load voltage within less than a cycle following a ±10\% disturbance in the grid input voltage. \textit{ii) Power factor correction:} by controlling the $q$-axis current of VSC2,  a PI loop compensates the reactive power drawn by the load and maintains unity power factor at the grid interface. \textit{iii) Frequency regulation:} the HDT modulates active power based on frequency deviations, following a droop curve derived from NESO data. Simulations show that the HDT can bid and track power setpoints equivalent to ±10\% of nominal voltage during frequency events. \textit{iv) Load phase balancing:} in the event of unbalanced three-phase loads, the HDT injects corrective current via VSC2 to balance the grid-side currents. The approach remains effective under 50\% imbalance in the RL load parameters. 
\textit{v) Simultaneous ancillary services:} the HDT simultaneously performs voltage regulation, power factor correction, frequency regulation, and load phase balancing, demonstrating coordinated multi-service delivery.

The HDT default parameters are defined in Table \ref{tab:parameters}.  

\begin{table}
\centering
\caption{Default simulation parameters.}
\label{tab:parameters}
\begingroup
\fontsize{6.7pt}{8pt}\selectfont
\begin{tabular}{llll}
\toprule
            Symbol &                                        Description &      Value & Unit \\
\midrule
           $\Delta t$ &                               Simulation time step &    $2 \times10^{-5}$ &    s \\
         $v_0$ &                               Nominal load voltage &      380.0 &    V \\
            $v_C^*$ &                          Reference DC-link voltage &     \textcolor{black}{ 2000.0} &    V \\
           $i_{2q}^*$ &                \makecell[l]{Reference $q$-axis current for VSC2 (except for \\ phase balancing and power factor correction)} &        0.0 &    A \\
           $i_{20}^*$ &            \makecell[l]{Reference zero-sequence current for VSC2 \\(except for phase balancing)} &        0.0 &    A \\
    $v_C(0)$ &                            Initial DC-link voltage &      1900.0 &    V \\
$I_i(0)$ &                            Initial currents $i=\{1, 2, 3\}$ &      0.0 &    A \\
$K_{p, 1}$ &       Proportional gain (load voltage controller) &       0.0 &    - \\
   $K_{i, 1}$ &           Integral gain (load voltage controller) &        1633.0 &    - \\
   $K_{p, 2}$ &       Proportional gain (shunt current controller) &       10.0 &    - \\
   $K_{i, 2}$ &           Integral gain (shunt current controller) &        8.0 &    - \\
$K_{p, 3}$ &       Proportional gain (DC-link voltage controller) &       0.14 &    - \\
   $K_{i, 3}$ &           Integral gain (DC-link voltage controller) &        0.01 &    - \\
      $K_{p, 4}$ &        Proportional gain (power factor correction) &        0.001 &    - \\
                $K_{i, 4}$ &            Integral gain (power factor correction) &         0.005 &    - \\
$K_{p, 5}$ &           Proportional gain (frequency regulation) &        0.01 &    - \\
      $K_{i, 5}$ &               Integral gain (frequency regulation) &       20.0 &    - \\
                   $K_{p, 6}$ &                 Proportional gain (phase balancing) &       50.0 &    - \\
                   $K_{i, 6}$ &                     Integral gain (phase balancing) &        1.0 &    - \\
  $L_p$ &                            Primary-side inductance &      0.795 &    H \\
$L_s$ &                          Secondary-side inductance &    $2 \times10^{-4}$ &    H \\
     $L_1$ &                                    VSC1 inductance &     0.0063 &    H \\
     $L_2$ &                                    VSC2 inductance &        0.2 &    H \\
   $R_p$ &                            Primary-side resistance &       50.0 &    $\Omega$ \\
$R_s$ &                          Secondary-side resistance &       0.01 &    $\Omega$ \\
     $R_1$ &                                    VSC1 resistance &      0.033 &    $\Omega$ \\
     $R_2$ &                                    VSC2 resistance &       0.01 &    $\Omega$ \\
      $C$ &                                DC-link capacitance &     0.0018 &    F \\
$L$ &                            Load inductance &     \textcolor{black}{ $8.3\times10^{-2}$} &    H \\
$R$ &                                    Load resistance &      \textcolor{black}{10.0}&    $\Omega$ \\
    $1/\alpha$&                  Grid-side transformer turns ratio &     1/66 &    - \\
  $\beta$ &                     Series transformer turns ratio &  18 &    - \\
              $V_{\text{in}, \text{pk}}$ &                                  Peak grid voltage &    25000.0 &    V \\
$I_{\text{pk}}$ &                                  Peak load current &  13.15 &    A \\
      $f_0$ &                                     Nominal grid frequency &       50.0 &   Hz \\
           $P_\text{max}$ &                         Maximum active power &      34458.6 &    W \\
           $P_\text{min}$ &                         Minimum active power &     24719.0 &    W \\
          $P_0$ &                        Nominal active power &     29160.0 &    W \\
        $\Delta f_\text{max}$ &                        Maximum frequency deviation &        0.2 &   Hz \\
                $\gamma$ &     Gain for $d$-axis correction in phase balancing &         10.0 &    - \\
\bottomrule
\end{tabular}
\endgroup
\end{table}
\normalsize
\subsection{Load voltage regulation}

In this scenario, the HDT regulates the load-side voltage under disturbed grid input conditions.  The voltage controller adjusts the duty cycle $D_1$ of VSC1 using the PI control law in equation~\eqref{eq:d1} to track a fixed load voltage reference of $v_0~=~380~\si{V}$. For ease of analysis, all voltage quantities are expressed in per-unit (pu), with the load voltage normalised by $v_0$ and the input voltage normalised by $V_{\text{in}, \text{pk}}$.

\begin{figure}[!ht]
\centering
  \includegraphics[width=0.9\columnwidth, trim={0cm, 0cm, 1cm, 0.5cm}, clip]{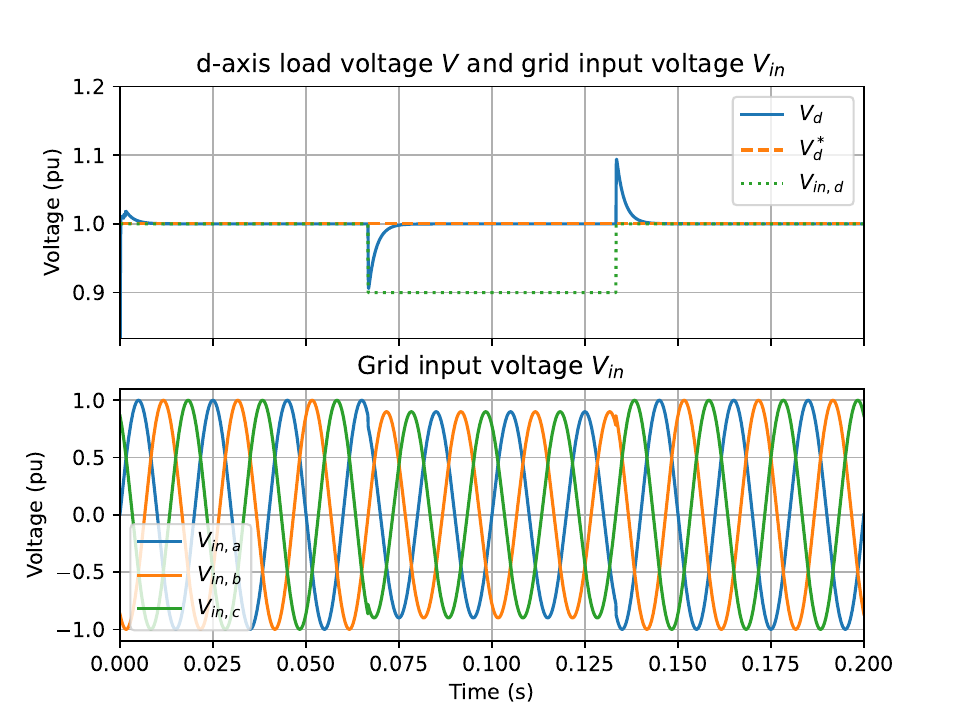}
  \caption{Voltage regulation under grid voltage disturbance. The HDT maintains the load voltage $V$ near its nominal value of $380~\si{V}$ (1 pu) despite variations in the input voltage $V_{\text{in}}$.}
  \label{fig:vreg}
\end{figure}

As shown in Figure~\ref{fig:vreg}, a ±10\% disturbance is introduced in the input grid voltage $V_{\text{in}}$  midway through the simulation (see dotted green curve representing the d-axis component of the grid voltage). The HDT compensates for this perturbation and maintains the load-side voltage $V_d$ close to its nominal reference of $1~\text{pu}$ (see plain blue and dashed orange curves). The controller exhibits a fast transient response with a settling time of approximately $0.01~\si{s}$ and no steady-state error. These results validate the effectiveness of the inner voltage regulation loop in rejecting grid-side voltage disturbances.

\subsection{Power factor correction}

In this case study, the HDT shunt converter compensates the reactive power drawn by the inductive load and maintains a unity power factor at the grid interface.  The DC-link voltage initial value is set to $v_C(0) = 2000 \si{V}$ for stability and all other parameters keep their default values as per Table~\ref{tab:parameters}.  To compensate the reactive power drawn by the load, the HDT modulates the duty cycle of VSC2 using the current controller in equation~\eqref{eq:d2}, such that the $q$-axis current tracks the reference $i_{2q}^*$ computed by the power factor correction loop in equation~\eqref{eq:pf_corr}.

\begin{figure}[!ht]
\centering
  \includegraphics[width=0.9\columnwidth, trim={0cm, 0cm, 1cm, 0.5cm}, clip]{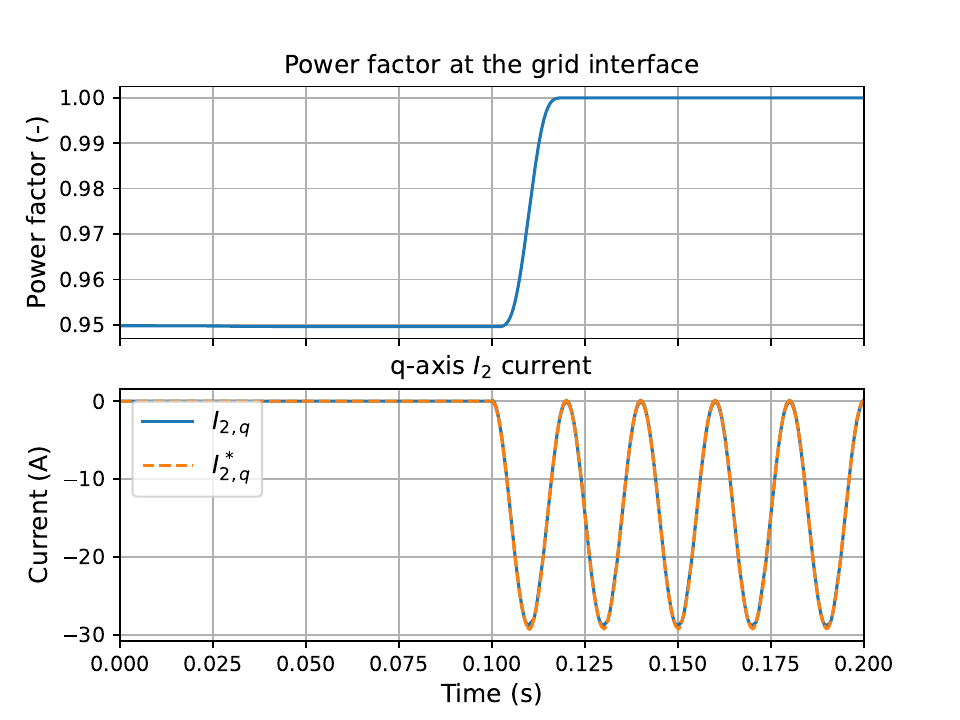}
  \caption{Top: Power factor at the HDT grid interface.  Bottom: $q$-axis component of the shunt converter current $i_{2q}$ and its reference $i_{2q}^*$.}
  \label{fig:pfcorr}
\end{figure}

Figure~\ref{fig:pfcorr} illustrates the system response when power factor correction is activated mid-simulation.  The top subplot shows the resulting power factor at the grid interface. Initially, the load operates with a lagging power factor of 0.95 ($i_{2q}^*~=~0~\si{A}$).  At $t = 0.1\,\si{s}$, the power factor correction controller in equation~\eqref{eq:pf_corr} is activated. The HDT increases the power factor to unity within roughly $0.015\,\si{s}$, demonstrating fast dynamic correction. The power factor then remains close to unity for the remainder of the simulation.  The bottom subplot shows that the $q$-axis current $i_{2q}$ accurately tracks the desired reference $i_{2q}^*$. 

\subsection{Load phase balancing}

In this scenario, we evaluate the ability of the HDT to correct unbalanced load conditions. The load impedance is modified to introduce asymmetries in both resistance and inductance:
\begin{gather}
L = \begin{bmatrix}
8.3\times10^{-2} & 1.6\times10^{-4} & 2.5\times10^{-4}\\
1.6\times10^{-4} & 8.3\times10^{-2} & 8.3\times10^{-5}\\
2.5\times10^{-4} & 8.3\times10^{-5} & 8.3\times10^{-2}
\end{bmatrix}\,\text{H},
\\
R = \begin{bmatrix}
10.0 & 0.1 & 0.2\\
0.1 & 10.0 & 0.15\\
0.2 & 0.15 & 5.0
\end{bmatrix}\,\Omega.
\end{gather}
To mitigate the resulting imbalance at the point of common coupling, the HDT adjusts the duty cycle of VSC2 based on the current control law defined in~\eqref{eq:d2}, using the reference shunt current computed in equations~\eqref{eq:balancing0}-\eqref{eq:balancing}. 

\begin{figure}[!ht]
\centering
  \includegraphics[width=0.9\columnwidth, trim={0cm, 0cm, 1cm, 0.5cm}, clip]{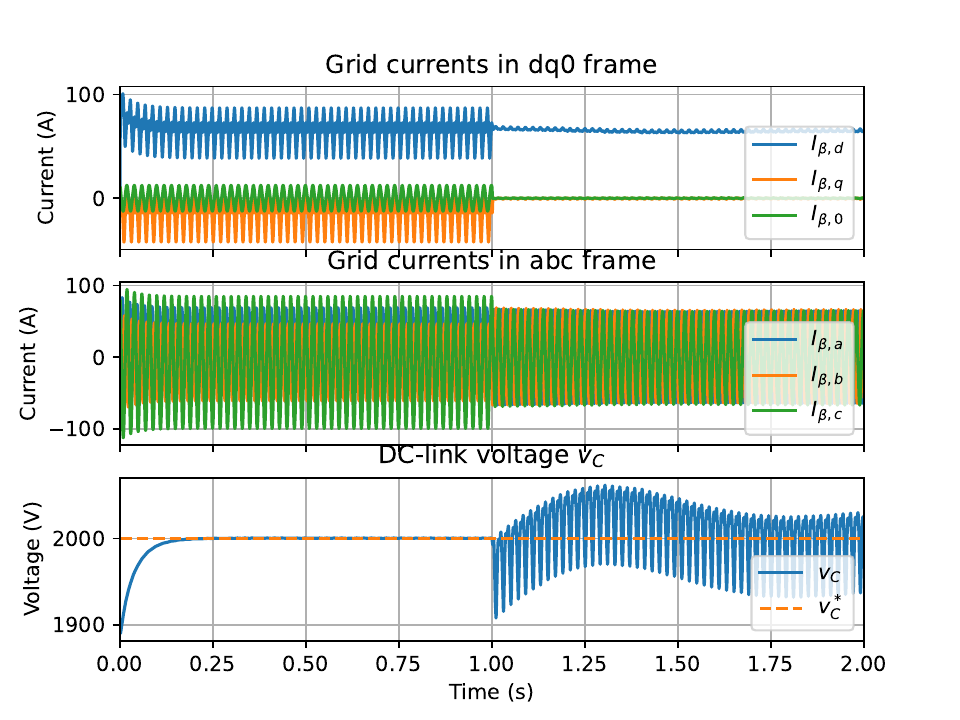}
  \caption{Grid current waveforms under unbalanced load conditions before and after activation of the load phase balancing controller. The middle subplot shows the phase currents in the \textit{abc} frame, highlighting the initial amplitude and phase mismatch. The top subplot displays the corresponding {dq0} components, where the $q$-axis and zero-sequence currents converge to zero after controller activation. The bottom subplot shows the DC-link voltage stabilised around its reference despite the corrective actions (with slight oscillations due to the trade-off between current balancing and energy storage stability).}
  \label{fig:balancing}
\end{figure}

Figure~\ref{fig:balancing} shows the grid current $I_\beta$ waveforms in both abc and dq0 frames, as well as the evolution of the DC-link voltage.  The balancing controller is activated at mid-simulation to highlight the corrective effect. Before activation, the currents are clearly unbalanced, with large deviations in amplitude and non-zero values in the $q$ and zero-sequence axes. Upon activation of the balancing controller, the HDT injects compensation shunt current that restores symmetry across phases and forces the dq0 components to converge toward their balanced targets. The DC-link voltage remains regulated around its reference $v_C^* = 2000\ \si{V}$ throughout the simulation, demonstrating that the load phase balancing operation does not compromise energy storage stability.  A $5\%$ degradation in DC-link voltage regulation is observed, which is expected due to the trade-off between current balancing and DC-link voltage stability objectives introduced in computing the desired $d$-axis component of the shunt current in equation \eqref{eq:balancing}. Note that this degradation can be reduced if $\gamma$ is increased in equation  \eqref{eq:balancing}, at the expense of load phase balancing. These results validate the capability of the HDT to enhance power quality under severe load asymmetries,  with $50\%$ resistance imbalance on phase~c.

\subsection{Frequency regulation}

This scenario demonstrates the HDT capability to provide dynamic frequency response by modulating the active power delivered to the load in real time according to grid frequency variations. The droop-based controller defined in equation~\eqref{eq:freq_P} maps deviations from the nominal frequency to a desired active power setpoint $P^*$, which is then converted into a corresponding voltage reference $v^*$ using equation~\eqref{eq:freq_v}. This reference is tracked by the inner-loop voltage controller defined in equation~\eqref{eq:d1}.

The frequency profile used in the simulation is obtained from historical NESO data \cite{NESOfrequency}, capturing a drop over a period of $T = 15\,\si{s}$. To improve stability, the DC-link capacitor voltage initial value is set to $v_C(0) = 2000\,\si{V}$, while all other parameters keep their default values as per Table~\ref{tab:parameters}.

\begin{figure}[!ht]
\centering
  \includegraphics[width=0.9\columnwidth, trim={0cm, 0cm, 1cm, 0.5cm}, clip]{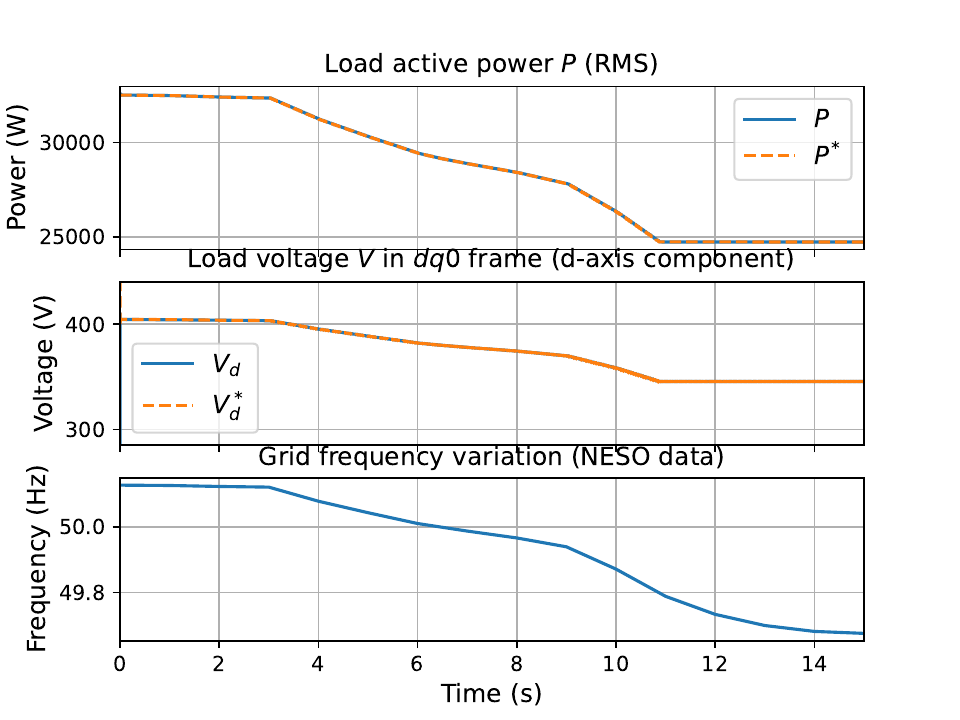}
  \caption{Dynamic frequency response. The bottom subplot shows the frequency profile. The droop-based controller translates this deviation into a desired active power trajectory (top subplot), which is achieved by adjusting the load voltage (middle subplot) through the inner-loop voltage control.}
  \label{fig:freq}
\end{figure}

Figure~\ref{fig:freq} presents the results. The bottom subplot shows the frequency profile, which decreases from $50.1\,\si{Hz}$ to $49.7\,\si{Hz}$. As expected, the frequency drop leads to a reduction in the desired active power setpoints and, consequently, in the load voltage reference—see the dashed orange curves in the top and middle subplots, respectively. The HDT responds by adjusting the voltage on the load side (plain blue curve in the middle subplot), resulting in an active power output that closely matches the desired trajectory (plain blue curve in the top subplot). This confirms the system ability to participate in frequency regulation services.

\subsection{Simultaneous ancillary service delivery}

\begin{figure}[!ht]
\centering
  \includegraphics[width=0.9\columnwidth, trim={0cm, 0cm, 0cm, 0.4cm}, clip]{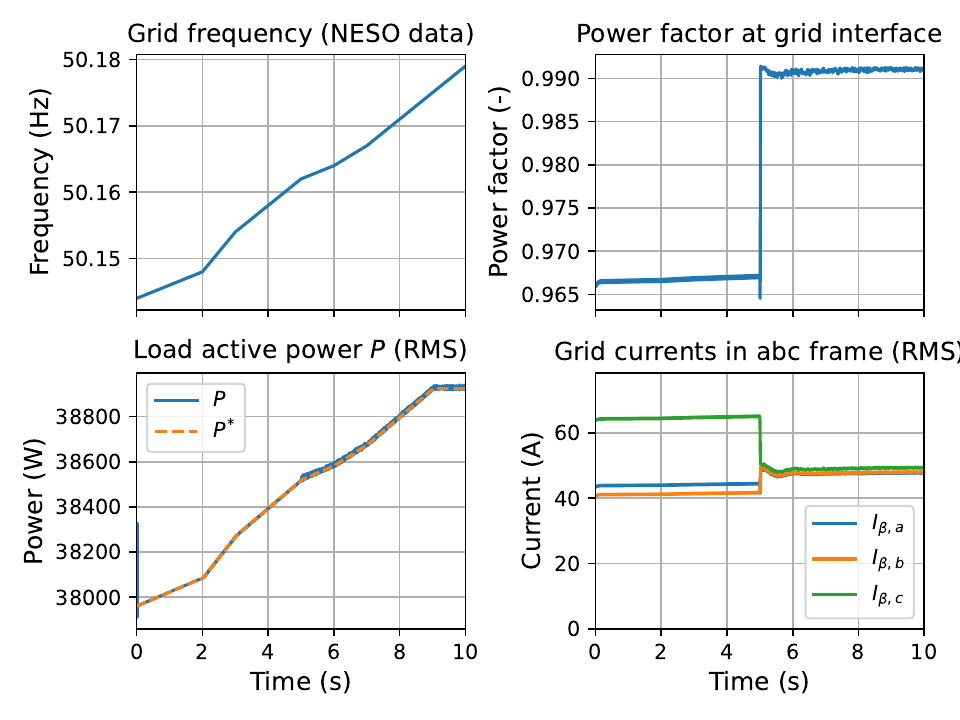}
  \caption{Simultaneous ancillary service delivery.  Top left: grid frequency profile. Bottom left: The active power at the load (solid blue) closely tracks the frequency-dependent reference \( P^* \) (dashed orange). Top right: power factor at the grid interface is maintained near unity (mid-simulation).   Bottom right: three-phase grid currents RMS values converge mid-simulation, showing balanced operation despite initial load asymmetries.}
  \label{fig:multi}
\end{figure}

To validate the versatility of the HDT, a final scenario considers the concurrent delivery of all four ancillary services. The system is subject to an unbalanced inductive load and grid frequency increase. The inner-loop controllers compute the duty cycles $D_1$, $D_2$ using~\eqref{eq:d1} and~\eqref{eq:d2}, driven by aggregated reference signals for voltage, active power, reactive power, and current balancing. To illustrate the attenuating effect of power factor correction and load phase balancing, the corresponding controllers are activated at mid-simulation. As shown in Figure~\ref{fig:multi}, the HDT simultaneously tracks the droop-based active power reference, improves the power factor to near unity, and attenuates current imbalances. This confirms that HDTs can provide simultaneous ancillary services with minimal tradeoff and high responsiveness.

\section{Conclusion}
\label{sec:conclusion}

This paper presented a dynamic model and control strategy based on PI regulators for hybrid distribution transformers (HDTs) aimed at delivering multiple ancillary services in distribution networks. We demonstrated the HDT ability to regulate voltage, correct power factor, support frequency response services through active power injection, and balance load currents across phases. Each functionality was validated independently as well as jointly in a multi-service scenario, showing that all objectives could be achieved simultaneously with fast response and minimal interference between control loops.  An analogy between the HDT in the series–shunt topology and UPQC devices was also highlighted, indicating that the methods developed in this work are broadly applicable and transferable to UPQC systems.

Future work will focus on enhancing the robustness of the control architecture against model uncertainties, disturbances and fault conditions (e.g.  single-phase failures) using stochastic model predictive control.  In addition, we plan to extend the framework to large-scale scenarios involving decentralised optimisation of multiple HDTs in order to maximise renewable energy integration onto the grid and provide frequency response support.

\bibliographystyle{IEEEtran}
\bibliography{biblio}

\end{document}